# Conferences with Internet Web-Casting as Binding Events in a Global Brain: Example Data From Complexity Digest

A. Das,
*Dept. of Mathematics*
*Jadavpur Univ., India*
dasatin@yahoo.co.in

C. Gershenson,
*Center Leo Apostel, Vrije Univ.Brussel,Belgium*
cgershen@vub.ac.be

G. Mayer-Kress,
*Dept. of Kinesiology*
*Penn State Univ., PA*
gxm21@psu.edu

P. Das
*Dept. of Mathematics*
*Jadavpur Univ., India*
prithadas01@yahoo.com

## Abstract

*There is likeness of the Internet to human brains which has led to the metaphor of the world-wide computer network as a `Global Brain'. We consider conferences as 'binding events' in the Global Brain that can lead to meta-cognitive structures on a global scale. One of the critical factors for that phenomenon to happen (similar to the biological brain) are the time-scales characteristic for the information exchange. In an electronic newsletter- the Complexity Digest (ComDig) we include webcasting of audio (mp3) and video (asf) files from international conferences in the weekly ComDig issues. Here we present the time variation of the weekly rate of accesses to the conference files. From those empirical data it appears that the characteristic time-scales related to access of web-casting files is of the order of a few weeks. This is at least an order of magnitude shorter than the characteristic time-scales of peer reviewed publications and conference proceedings. We predict that this observation will have profound implications on the nature of future conference proceedings, presumably in electronic form.*

## 1. Introduction

### 1.1 Brain

Neurons are the building blocks of brain. Of the order of ten billion neurons and their interconnections in our brain make human neural networks one of the most complex networks. Real neurons also exhibit multiple and often multistep processes. One neuron accepts the output of other neuron(s) as input - synaptic weights at the junction of the communicating neurons determine the strength of connection. These synaptic weights are continuously modified by our physical and psychological activities [1].

### 1.2 Global brain

There is likeness of the Internet to human brains. For example, the organization of information on the world wide web (WWW) with hypertext links closely resembles the associative connections formed by neurons in the brain [2,3,4]. The Internet network encompasses the entire globe, holding the possibility for linking all humans with a means of virtually instant interaction. The Internet has a decentralized, self-organizing nature, where information will travel through whatever route available.

Particularly, the web is the hypermedia interface to the information on the Internet allowing hyperlinks to other documents. Thus, the web makes it possible to seamlessly integrate documents that are distributed over the entire planet. In that respect, this hypermedia architecture is similar to the one of our brain, where concepts are connected by associations, and the corresponding assemblies of neurons by synapses.

Moreover, the different 'nodes' of the digital network - controlled by computers allow sophisticated processing of the collected data, reinforcing the similarity between the network and the brain. This has led to the metaphor of the world-wide computer network as a `*global brain*' [2]. The parallels between biological brains and the global Internet have been discussed in some details in the literature [2 and references therein].

### 1.3 Binding problem

One central question in cognitive neural science is the binding problem. In general, binding involves grouping features into objects, that is, "binding" refers to the association of information for some perceptual unity such as in the visual system. This is a crucial issue in the learning process in the human brain. Researchers know that information about what an object looks like and where the object is are stored in two separate sections of the brain and analyzed in separate parts of the visual system. The question is where and how does this information come together? This problem is better known as the 'binding problem' [5]. In experiments with monkeys [6], it was found that "neurons in the prefrontal cortex had activity that exquisitely represented and integrated both 'what' and 'where.' This provides a first clue on how this information, and perhaps even more diverse information, comes together in the brain. While early results seem to flag the prefrontal cortex as a prime binding location, it also is possible that the information could be coming together elsewhere in the brain first.

### 1.4 Role of conferences

The formation of cell-assemblies in the biological brain is believed to be associated with cognitive events and feature binding in perception and learning. In the context of a global brain we interpret gatherings of intelligent agents with the objective to communicate intensively on a common topic as an analogous event [7]. We know from biological brains that the formation of cell assemblies also is accompanied by gamma band ("40Hz"), synchronized electrical activity of participating neurons. Thereby characteristic time-scales of biological brains are established. We know that in human communication there exist similar "universal" time-scales that facilitate constructive interaction and the emergence of collective, self-organized behavior. One of the shortest time-scales is that of synchronous interaction (e.g. 300ms in conversation) but other time-scales determined by biological factors can be of similar importance. Current mega-conferences with the order of $10^4$ participants push the limits of the concept of "face-to-face" interactions among participants. At the same time we currently witness the advent of modern electronic communication tools that could push the envelope of meaningful interactions in big conferences all the way to scales relevant to global brain dimensions ($10^{10}$). We consider conferences as 'binding events' in the Global Brain that can lead to meta-cognitive structures on a global scale. One of the critical factors for that phenomenon to happen (similar to the biological brain) are the time-scales and efficiency of the information exchange among the on-site presenters and local and global audience. For conferences our goal is to make digital recordings of the event available on the Internet in timescales of less than one week.

### 1.5 Time Scales

Characteristic Time Scales Related to a Global Brain are the following [8,9]:
Synchronous (talk, video-conference): ~300ms  ("Instantaneous")
Information Retrieval: min (Library)
Attention: hour (Lecture)
Asynchronous (e-mail): day ("Sleep over it")
Publication: week, month (Peer Review)
Half-Life of Collective Activity (active use/update of website), year

### 1.6 Aim of this paper

We will present the analysis of real-world data to show what we have already achieved in this regard and estimate the binding factor. We shall see that the time scale of the binding factor in the global brain is of the order of few weeks.

## 2. Data

The Complexity Digest (ComDig) is an edited electronic newsletter in the field of complexity; issues are published on weekly basis since 1999. The mirrored URLs are at www.comdig.org , www.comdig.de and http://www.phil.pku.edu.cn/resguide/comdig/
Keeping with the scope of greater global access to information over the Internet, we report International conferences/Seminars in video/audio formats. These files, as well as the issues of ComDig are free of charge thanks to generous private support. In order to analyze the nature of access to the posted conference files, we collected related data from our server for 10 months (Mar. 2002 to Feb. 2003). The web statistics is grouped on the basis of access for six days after the date of publication of an issue. Here "access" is defined by the analysis software, provided by the Internet service provider. Data contains total number of accesses during any week, file-wise access numbers, etc.

## 3. Analysis

Fig. 1 shows the total number of accesses to www.comdig2.de during the observation period mentioned above. www.comdig2.de is the site that hosts ComDig conference webcasts. Thus the access figures do not reflect the readership of the newsletter itself.

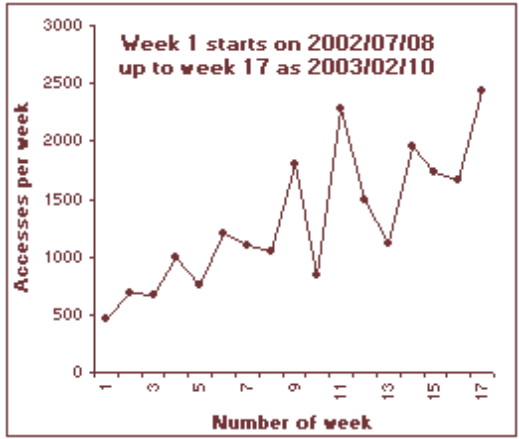

**Figure 1. Overall number of accesses per week to www.comdig2.de pages. Note that this is not one of the official archive sites of ComDig.**

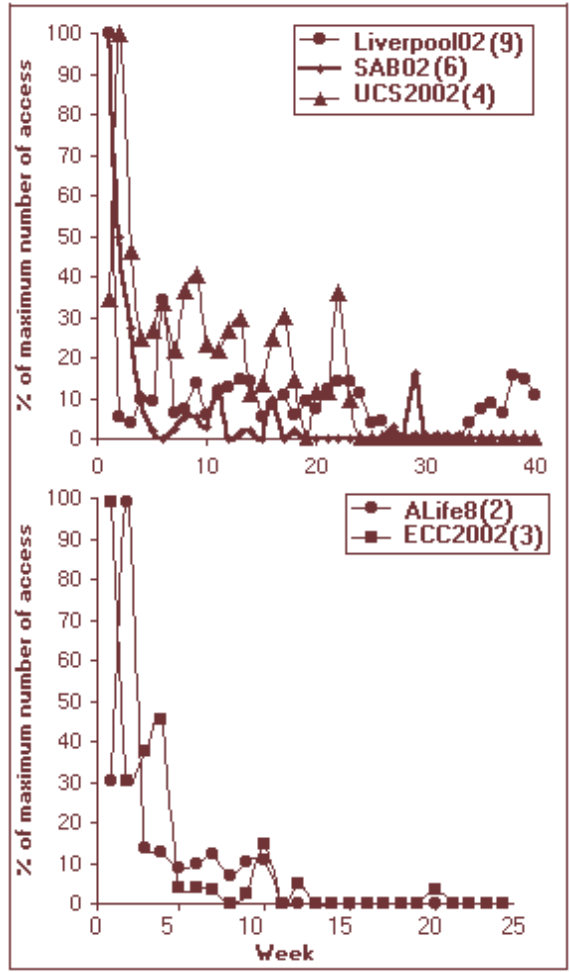

**Figure 2a (above) & 2b(below). Showing sharp fall in number of accesses per week. Inset: curve type- conference name (difference between last day of conference & the day posted in ComDig)**

From Fig. 2 it is evident that for some conference (those shown in Fig. 2b) accesses to the conference media files decays to less than 50% of the maximum rate already after one week. The asymptotic rate of less than 10% is reached after about five weeks. Note that for any conference, we have plotted percent values of maximum accesses recorded.

## 4. Nature of Access

Fig. 3 shows the actual accesses per week for the Alife8 conference. We can clearly see that the asymptotic rate is reached two weeks after the conference.

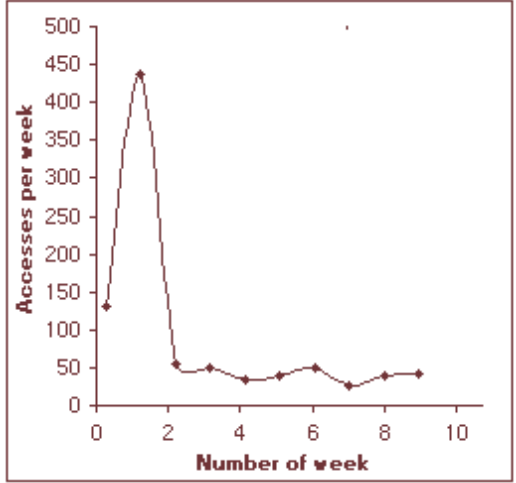

**Figure 3. Accesses per week for ALife8 files**

In Fig. 4, we are showing axonal pulse train of a single neuron or a small group of neurons, yielding a post-stimulus histogram (PSTH) time- smoothed by drawing a curve- as given in Fig. 4. Small triangles show PSTH of a group of mutually excitatory neurons in the olfactory bulb for an excited stimulus (black dot).

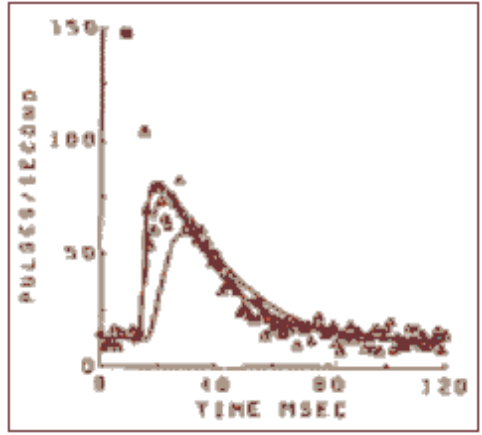

**Figure 4. Axonal pulse train of neuron(s). Original Fig. by Freeman [1] and reproduced with permission.**

Comparison of Figs. 3 & 4 shows again that there is likeliness to the actual brain and global brain metaphor. Also estimation from Fig. 4 shows that conferences serve as binding factor to the global brain with times scales of few weeks.

## 5. Different types of media files

We posted both video and audio formats of some presentations and will like to compare the access of these files. The files are in video (.asf) and audio (.mp3) format. Note that the videos are of 5 mins (approximately) while audio (mp3) files are of 30-60 mins. duration. Preference of video files may also indicate that people prefer also to spend less time sitting before a computer to listen these files.

This ratio in favor of asf can change rapidly if two parameters change: (1) The Internet access speed (mp3 files are large and require longer download times), (2) The market penetration of portable mp3 music players. Currently those players are very common among teenagers who are not the typical audience of ComDig. When this technology is familiar to the adult population we expect that the advantages of being able to listen to presentations without having to sit in front of a computer will reverse that ratio.

**Table1: Showing comparison of different types of media files**

| File name | philanderson | | stevehassan | |
|---|---|---|---|---|
| Format | .asf | .mp3 | .asf | .mp3 |
| Total accesses | 192 | 28 | 477 | 45 |

## 6. Discussion

Apart from estimating the characteristic time scales related to interest in conference presentations to be in the order of few weeks, we can also conclude that video formats are more currently the more popular choice for the users. The characteristic time-scales of web-casting of conference presentations (as well as those related to Internet publication of presentation visuals) therefore can be seen to be at least an order of magnitude faster than traditional publication time-scales for paper proceedings. We expect that these observations will have significant impact on future documentation of conferences and publication of conference proceedings.

### Acknowledgement

This research is supported by Complexity Digest.

## 7. References

[1] W. J. Freeman, "Tutorial On Neurobiology: From Single Neurons To Brain Chaos", *Int. J. of Bifur. & Chaos*, vol. 2, no 3, World Science, Singapore, 1992, pp.451-482.

[2] G. Mayer-Kress, and C. Barczys, "The Global Brain As An Emergent Structure From The Worldwide Computing Network, And Its Implications For Modeling", *The Information Society*, vol. 11, no 1, Jan-Mar 1995, pp.1-27.

[3] F. Heylighen, "The Social Superorganism And Its Global Brain", 2000/03/23 at http://pespmc1.vub.ac.be/SUPORGLI.html

[4] F. Heylighen, and J. Bollen, "The World-Wide Web As A Super-Brain: From Metaphor To Model" in: R. Trappl (ed.) Cybernetics and Systems, World Science, Singapore, 1996 at http://www.c3.lanl.gov/~rocha/embrob/smith.html

[5] L. B. Smith, "The Body, Binding And Learning Words", IISREEC, Lisbon, Portugal, 2002/11/12-15.

[6] "Miller Probes Individual Neurons For Clues To Learning And Memory", *MIT Tech* Talk, 2000/05/31 at http://web.mit.edu/newsoffice/tt/2000/may31/miller.html

[7] G. Mayer-Kress, "Time-Scales And The Historical Role Of Conferences For The Emergence Of Global Brains", in From Intelligent Networks to Global Brains, Evolutionary Social Organization through Knowledge Technology, Brussels, Belgium, 2001/07/3-5.

[8] K. Newell, Y. T. Liu, and G. Mayer-Kress, "Time Scales In Motor Learning And Development", *Psychological Review*, vol. 108, no.1, 2001, pp:57-82.

[9] G. Mayer-Kress, "Localized Measures for Non-Stationary Time-Series of Physiological Data", *Integrative Physiological and Behavioral Sc.*, vol. 29, no.3, July-Sept. 1994, pp.205-210